\documentclass{article}

\usepackage{microtype}
\usepackage{graphicx}
\usepackage{subfigure}
\usepackage{booktabs} 
\usepackage{verbatim}
\usepackage{enumitem}
\usepackage[hyphens]{url}
\usepackage[breaklinks=true]{hyperref}
\usepackage{array}
\usepackage{booktabs}
\usepackage{multirow}
\usepackage{tabularx}
\usepackage{longtable}
\usepackage{array}
\usepackage{booktabs}
\usepackage{multirow}
\usepackage[table]{xcolor}
\usepackage{natbib}

\usepackage{rotating}

\usepackage{hyperref}

\usepackage[accepted]{icml2025}

\usepackage{amsmath}
\usepackage{amssymb}
\usepackage{mathtools}
\usepackage{amsthm}

\usepackage[capitalize,noabbrev]{cleveref}

\theoremstyle{plain}

\theoremstyle{definition}

\theoremstyle{remark}

\usepackage[textsize=tiny]{todonotes}

\icmltitlerunning{Societal Capacity Assessment Framework}

\begin{document}

\twocolumn[
\icmltitle{Societal Capacity Assessment Framework:\\ Measuring Resilience to Inform Advanced AI Risk Management
}



\icmlsetsymbol{equal}{*}

\begin{icmlauthorlist}
\icmlauthor{Milan Gandhi}{yyy,equal}
\icmlauthor{Peter Cihon}{comp,equal}
\icmlauthor{Owen Larter}{xxx}
\icmlauthor{Rebecca Anselmetti}{zzz}
\end{icmlauthorlist}

\icmlaffiliation{yyy}{Oxford Martin AI Governance Initiative}
\icmlaffiliation{comp}{contribution while at GitHub}
\icmlaffiliation{xxx}{contribution while at Microsoft}
\icmlaffiliation{zzz}{UK AI Security Institute}

\icmlcorrespondingauthor{Milan Gandhi}{milan.m.gandhi@outlook.com}

\icmlkeywords{Machine Learning, ICML}

\vskip 0.3in
]



\printAffiliationsAndNotice{The views expressed in this paper are those of the authors alone and do not necessarily reflect the official policy or position of their employers. * equal contribution} 

\begin{abstract}
Risk assessments for advanced AI systems require evaluating both the models themselves and their deployment contexts. We introduce the Societal Capacity Assessment Framework (SCAF), an indicators-based approach to measuring a society's vulnerability, coping capacity, and adaptive capacity in response to AI-related risks. SCAF adapts established resilience analysis methodologies to AI, enabling organisations to ground risk management in insights about country-level deployment conditions. It can also support stakeholders in identifying opportunities to strengthen societal preparedness for emerging AI capabilities. By bridging disparate literatures and the ``context gap'' in AI evaluation, SCAF promotes more holistic risk assessment and governance as advanced AI systems proliferate globally.

\end{abstract}

\vspace{-20pt}
\section{Introduction}
\label{submission}

Risk is often defined as encompassing two dimensions: (1) the \textit{likelihood} of a hazard occurring and (2) the \textit{severity} of its impact. The risks of frontier general-purpose AI systems (hereinafter ``advanced AI''), including that they may be misused as part of cyber, chemical, and biological attacks \cite{metr2025common,bengio2025international}, are shaped not only by the capabilities of AI models but also by external variables \cite{bengio2025international,nist2025updated,solaiman2023evaluating}. These include how models are deployed as part of end-to-end systems, the scale of deployments, and interactions between AI capabilities, individual users and broader economic, environmental, and societal systems \cite{weidinger2023sociotechnical,berman2024troubling}. This helps to explain why governance tools, such as pre-deployment model capability evaluations, though central to safety frameworks, have limited predictive power in assessing risk \cite{burden2024evaluating,weidinger2023sociotechnical}. Enhancing the external or ‘ecological’ validity of AI evaluations represents an important unresolved challenge in AI governance research \cite{reuel2024open,ukaisi2025,bengio2025singaporeconsensusglobalai}. 

Responding to this challenge, we introduce the Societal Capacity Assessment Framework (SCAF). SCAF provides a method to structure assessments of a social unit, such as a country, based on its vulnerability and resilience to advanced AI risks. It draws from indicators-based approaches used for assessing societal and systems' resilience in other settings, as well as recent contributions to the AI governance literature. SCAF complements existing approaches for evaluating advanced AI capabilities with consideration of the characteristics, resources, and policies of communities exposed to AI risks.

SCAF can be used by AI developers, governments, and other stakeholders to inform, \textit{inter alia}:
\begin{itemize}[itemsep=5pt, topsep=0pt, parsep=0pt]
\item Pre-deployment risk assessments by incorporating analysis of deployment contexts into risk calculations that are today heavily and narrowly reliant on model evaluations and limited threat modeling. 
\item Adaptation and societal resilience planning, by exploring priority areas for policy interventions and investments to strengthen societal preparedness for advanced AI (e.g., \citeauthor{toner2025}, \citeyear{toner2025}).
\end{itemize}

We highlight literature informing SCAF (\hyperref[sec:lit]{Section 2}), linking AI governance scholarship with approaches to resilience analysis from other fields. We then introduce SCAF (\hyperref[sec:SCAF]{Section 3}) and discuss its potential use cases and avenues for improving the framework, including directions for future research (\hyperref[sec:discussion]{Section 4}). \hyperref[appendixa]{Appendix A} presents a prototype of SCAF and illustrates its possible application to Australia.

\section{Selected Literature}
\label{sec:lit}

\subsection{AI Governance and Sociotechnical Evaluations}
Weidinger et al. (\citeyear{weidinger2023sociotechnical}) propose a sociotechnical approach to evaluating risks of generative AI systems, observing the gap that arises when ‘AI system safety is evaluated only with regard to technical components of an AI system…’ They recommend complementing technical evaluations of AI capabilities with assessments of human-system interaction and ‘the impact of an AI system on the broader systems in which it is embedded, such as society, the economy, and the natural environment’. Reviewing existing practices, they identify a ‘context gap’, noting that such broader evaluations remain ‘rare’. Berman et al. (\citeyear{berman2024troubling}) likewise argue that assessing the societal impacts of generative AI requires ‘a model of how social harms emerge from interactions between [generative AI], people, and societal structures’, emphasising the need to make such models explicit. 

AI developers also recognise the ‘context gap’. For example, Microsoft's Frontier Governance Framework (\citeyear{microsoft2025frontier}) and Google DeepMind's holistic approach to evaluating advanced AI models \cite{weidinger2024holistic} both acknowledge that AI is a sociotechnical system and harms emerge from the technology's interaction with individual, societal, and environmental conditions. 

Solaiman et al. (\citeyear{solaiman2023evaluating}) propose a ‘framework for social impact of generative AI systems’, noting that generative AI models ‘have no specific predefined application, sector, or use case, making them notoriously difficult to evaluate’. Their framework defines and organises social impact categories at both the base technology level (e.g., the AI model) and the ‘people and society’ level. Within the ‘people and society’ level, the authors identify  multiple evaluation methods, including system-level experiments, post-deployment monitoring measures, and ecosystem monitoring for particular categories like economic and labour impacts. 

Bernardi et al. (\citeyear{bernardi2024societal}) advocate for societal adaptation to advanced AI, arguing that focusing solely on modifying AI systems’ potentially harmful capabilities (the raison d’\^etre of existing frontier safety frameworks) will become less effective over time as training costs fall and more powerful systems are deployed without safeguards. Kembery (\citeyear{kembery2024towards}) has similarly observed that ‘a more gradual diffusion of risky [AI] capabilities’, rather than a single point of failure, may become the governance paradigm for AI, complicating regulatory visibility and enforcement. Therefore, Bernardi et al. (\citeyear{bernardi2024societal}) propose a framework to ‘guide thinking about’ societal adaptation to advanced AI, outlining the ‘structure of a causal chain leading to negative impacts from AI’ in specific threats and providing ‘a categorisation of interventions' to reduce instances of such threats.

SCAF serves to narrow the gap identified by Weidinger et al. (\citeyear{weidinger2023sociotechnical}) at the broader systems level, presenting a framework to assess a society's vulnerability and resilience to the risks of advanced AI. We proceed from a similar perspective to Bernardi et al. (\citeyear{bernardi2024societal}) in that we look beyond risk mitigation at the technology level to consider societal factors that affect resilience to those risks. Bernardi et al. (\citeyear{bernardi2024societal}), however, focus on defining adaptation to AI through interventions along the causal chain of harm. Solaiman et al. (\citeyear{solaiman2023evaluating}) outline domain-specific methods for evaluating the social impacts of generative AI. Our point of difference is that we seek a pragmatic, standardisable, and data-driven way to assess overall societal vulnerability and resilience to advanced AI risks. This leads us to adapt resilience-analysis methodologies from other domains to the advanced AI context. 

\subsection{The Indicators-Based Approach to Analysing Societal Resilience}
\label{sec:2.2}
Defining methods for analysing the resilience of societies and complex systems has long been a priority for practitioners and scholars in engineering, social science, and public policy. The OECD has published guidelines for resilience systems analysis (\citeyear{oecd2014guidelines}) and reported on how 20 countries conduct their own national risk assessments (\citeyear{oecd2017nra}). The EU Commission supported research to develop a methodology for evaluating critical infrastructure resilience (e.g., \citeauthor{lange2017framework}, \citeyear{lange2017framework}). NIST reviewed indicators of community resilience to natural disasters \cite{gerst2024review} and has published an inventory of indicators and assessment frameworks \cite{dillard2021inventory}. The UK’s National Risk Register and Resilience Action Plan outline a dynamic risk assessment process to assess national resilience to the most serious threats \cite{UKriskregister}. The UN Global Risks Report synthesises risks viewed as most important by stakeholders in government, the private sector, civil society, and academia, and assesses the preparedness of multilateral institutions based on average scores for risk identification, reduction, and mitigation \cite{UNrisks}. 

SCAF draws on this work, and especially approaches to ex ante assessments, i.e., examining resilience before a shock occurs, in contexts where there may be little to no data on past responses, but a structured forecast of potential outcomes is still essential. Forecasting the impacts of advanced AI exemplifies such a context, where ‘policymakers will often have to weigh potential benefits and risks of imminent AI advancements without having a large body of scientific evidence available’ \cite{bengio2025international}.

Measuring the resilience of societies and other complex systems is widely recognised as challenging \cite{gerst2024review}, with many potential variables bearing upon the outcome of resilience \cite{alheib2016report,rosenqvist2018isra,saja2019critical}. Determining how to weight variables and map causal interdependencies is complicated, particularly when data about past shocks is sparse \cite{rosenqvist2018isra,saja2019critical}. Moreover, resilience has many meanings, simultaneously understood as an attribute (or capacity), a process, and an outcome that reflects successful adaptation to, or recovery from, adversity \cite{pfefferbaum2013communities,copeland2020measuring}.

Not without controversy, ‘societal resilience’ (sometimes interchangeable with ‘social’ or ‘community’ resilience) is defined as ‘the ability of social groups or communities to cope with external stresses and disturbances’ \cite{rosenqvist2018isra}. Resilience is sometimes decomposed into distinct yet interlinked ‘capacities’ \cite{copeland2020measuring,rosenqvist2018isra,parsons2016top}. These include: \textit{coping capacity}, the short-term ability to ‘respond, absorb, and recover from a disruptive event’; \textit{adaptive capacity}, longer-term ability to ‘plan for and adjust to future challenges’; and \textit{transformative capacity}, transforming ‘the stability landscape’ or creating ‘new, better pathways’ for the system as a whole, which can involve fundamental institutional or structural change. 

There is a tension between the coping capacity – understood as preserving an existing system’s structure and identity – and adaptation and transformation, which can call for innovation and change rather than a return to the status quo \cite{pearson2012societal,oconnell2015resilience,copeland2020measuring}. This tension is relevant to advanced AI, which not only poses risks but also offers transformative opportunities, including safeguards against the very threats it may contribute to. For example, in adapting to the cyber risks posed by advanced AI, Bernardi et al. (\citeyear{bernardi2024societal}) recommend societies adopt ‘AI-enhanced cyber defence’. Drawing from development studies, enhancing resilience is sometimes theorised to depend on strengthening categories of assets or ‘capitals’, including human, social, financial, natural, physical, and political capital \cite{oecd2014guidelines}. Governance and public policies shape how these capitals are mobilised, distributed, and converted into resilience outcomes \cite{Carney1999}.

Holding resilience considerations constant, \textit{vulnerability} refers to a society's ‘propensity to be adversely affected' \cite{lecina2024resilience} and to the underlying conditions that ‘increase [society's] susceptibility... to the impacts of hazards' \cite{undrr2017vulnerability}.   

In bridging resilience theory and practice, numerous frameworks adopt an indicators-based approach \cite{norris2008community,oecd2014guidelines,parsons2016top,lange2017framework,copeland2020measuring,gerst2024review}. This involves two steps. First, \textit{concept definition}: defining the outcome of interest (e.g., societal resilience to forecasted shocks) and identifying latent constructs theorised to influence this outcome – such as the capacities and capitals introduced above. These constructs represent the explanatory mechanisms rather than directly observable variables. Their selection and definition is informed by hypothesising causal chains leading to shocks, which others are working to illustrate in the advanced AI context \cite{bengio2025international,bernardi2024societal}. Second, \textit{designing a measurement framework}: selecting directly measurable indicators that serve as empirical proxies for each construct.

\section{Resilience Indicators for Advanced AI}
\label{sec:SCAF}
Below we present SCAF, an indicators-based approach to measuring a society's vulnerability, coping capacity, and adaptive capacity in response to AI-related risks.

\begin{table*}[htbp]
\centering
\caption{Summary Presentation of SCAF}
\footnotesize 

\begin{tabularx}{\textwidth}{|>{\raggedright\arraybackslash}p{0.15\textwidth}|>{\raggedright\arraybackslash}X|>{\raggedright\arraybackslash}X|}
\hline
 & \textbf{Cyber risks} & \textbf{Chemical and biological (CB) risks} \\
\hline
\textbf{Vulnerability} & 
Indicators could measure national reliance on digital services and the quantity and severity of cyber attacks.
& 
Indicators could measure population health, healthcare system (baseline) capacity, population flows and density, the presence of extremist or terrorist groups, and the availability of CB weapons materials, expertise, and equipment.
\\[10pt] 
\cline{2-3}
& \multicolumn{2}{>{\raggedright\arraybackslash}p{0.80\textwidth}|}{
\textbf{Cross-cutting vulnerability indicators} could measure economic development (e.g., GNI per capita, GDP, population below \$2.15 per day), state capacity (e.g., tax revenue as a share of GDP, government territorial control, impartiality of public services), and democracy and peace.
} \\
\hline
\textbf{Coping Capacity} & 
Indicators could measure cyber incident tracking and response, including in critical infrastructure, government, and other sectors; the availability of cybersecurity professionals (e.g., proxied by GitHub users in a jurisdiction); and patch adoption rates.
& 
Indicators could measure health emergency response plans, the capacity of emergency response and crisis management services, extent of specialised CB training for responders and medics, availability of CB specialists, and medical countermeasure stockpiles.
\\
\hline
\textbf{Adaptive Capacity} & 
Indicators could measure the medium to long-term cybersecurity talent pipeline, secure-by-design initiatives, adoption of cybersecurity best practices by local websites and open-source developers, local implementation of NIS-2 (or equivalent) standards, critical infrastructure protection, AI literacy, cyber hygiene, and cross-sector security standards.

\vspace{4pt}

Indicators for \textit{transformative capacity} could measure the existence and extent of active cyber defence, AI-driven threat prediction, and continuous upskilling programs.
& 
Indicators could measure medium- to long-term health emergency preparedness planning, CB procurement and monitoring (e.g., synthetic nucleic acid procurement screening), CB equipment standards, threat sharing, infrastructure protection, and citizen preparedness. 

\vspace{4pt}

Indicators for \textit{transformative capacity} could measure AI-enabled biological and disaster monitoring, detection and forecasting; and AI-enabled patching for expanded synthetic nucleic acid procurement screening.
\\
\hline
\end{tabularx}
\label{tab:scaf-summary}
\end{table*}

\subsection{Concept Definition}
In defining the concepts that underpin SCAF, we balanced competing aims: fidelity to the societal resilience literature (\hyperref[sec:2.2]{2.2}), conceptual clarity \cite{gerring1999concept}, and a desire to minimise overlap between the concepts to preserve their analytical value as part of a measurement framework. SCAF organises indicators into three categories: \textit{vulnerability}, measuring structural and background conditions that predispose a society to harm; \textit{coping capacity}, measuring resources and systems that enable response and recovery in the immediate aftermath of a shock; and \textit{adaptive capacity}, measuring public policies and other interventions aiming to durably reshape the safety landscape. 

\textbf{Vulnerability} measures structural and background conditions that increase a society’s susceptibility to the risk in question \cite{undrr2017vulnerability}. Proxy indicators are readily available for potentially relevant conditions at the country-level such as development status, population health, and the quality of infrastructure and public services. A limitation of our vulnerability definition is potential overlap with coping capacity. To avoid this, we considered defining vulnerability as the harm that would result in the absence of a meaningful societal response. However, we decided that this counterfactual approach, while analytically precise, would be difficult to operationalise.

\textbf{Coping capacity} refers to resources and systems in place to absorb, respond to, and recover from an adverse shock in its immediate aftermath, such as emergency management and response services. For example, in the case of chemical or biological attacks, relevant indicators could include the availability of health emergency preparedness and response services, as well as the depth of medical countermeasure stockpiles.

\textbf{Adaptive capacity} encompasses public policies and other interventions aimed at mitigating, planning for, and adjusting to risks over the medium- to long-term, including measures to reduce the likelihood of adverse impacts and their severity. Certain adaptive interventions identified by Bernardi et al. (\citeyear{bernardi2024societal}) are relevant under this pillar. For now, we do not treat transformative capacity (see \hyperref[sec:2.2]{2.2}) as a distinct category. Rather, we also regard as part of this pillar adaptive interventions that reshape societal systems in lasting and systemic ways, including through the use of innovation and emerging technologies.

\subsection{Measurement Framework}
To develop a measurement framework for SCAF, each independent variable is approximated by indicators that capture a region’s vulnerability, coping capacity, and adaptive capacity, organised according to risk category. In terms of data sources, SCAF could, for example, utilise self-assessment (e.g., questionnaire completed by local government officials), expert assessment (e.g., questionnaire completed by regional subject matter expert), quantitative data (e.g., numerical scores based on proxy variables for each indicator), or a blend. SCAF results could be framed quantitatively or qualitatively, such as through aggregate scores or qualitative labels like “high readiness” \cite{yang2023indicator}. 

We illustrate an example implementation of SCAF (see  \hyperref[tab:scaf-summary]{Table 1} and \hyperref[appendixa]{Appendix A}) using two advanced AI risk categories: cyber offence, and biological and chemical attacks. These are highlighted by the 2025 International AI Safety Report \cite{bengio2025international} and have been operationalised by AI developers through capability thresholds employed in their frontier safety frameworks \cite{metr2025common}. As we outline in \hyperref[sec:discussion]{Section 4}, SCAF's risk coverage should be broadened. Similarly, our selection of indicators is indicative only, and future work should make explicit the conceptual and empirical grounds for their selection.

\subsection{Assumptions and Limitations}
A framework such as SCAF is not fully neutral. Its design embeds assumptions that shape its use and the decisions it informs \cite{copeland2020measuring}. Firstly, like other indicator-based approaches, SCAF assumes that identifying present variables may help prevent undesirable future outcomes \cite{copeland2020measuring}.
Secondly, SCAF assumes the utility of assessment at the country-level, but its conceptual framework can also be adapted to other units of analysis, including regions, cities, sectors. Thirdly, SCAF is a point-in-time assessment though social mechanisms shaping resilience are inherently dynamic and continuous in reality  \cite{saja2019critical}. Future iterations of SCAF could explore integrating time-series data.

SCAF aims to translate abstract and complex societal outcomes, such as coping with and adapting to the impacts of advanced AI risks, into a concrete and measurable framework. This process of translation is inherently limited due to the wide range of relevant variables, their likely redundancy over time, the difficulty of determining their relative importance and observing interdependencies, data limitations, and subjective elements of resilience (such as communal attitudes to risk), which vary across communities and cultures \cite{copeland2020measuring}. Further, evidence about frontier AI risks and the effectiveness of technical and public policy mitigations is still emerging \cite{bengio2025international}. Given these limitations, SCAF should be treated as a preliminary input to broader risk assessment and adaptation planning for advanced AI, and ideally developed and piloted within representative governance structures (discussed further in \hyperref[sec:discussion]{Section 4}). 

\subsection{Illustrative Case: Australia}
Using publicly available data, we complete a preliminary SCAF for Australia. Results of this exercise, along with expanded discussion, are presented in \hyperref[appendixa]{Appendix A}. Aggregate scoring or labels were not attempted for this exploratory exercise; instead, it was undertaken to provide some insights into practical challenges, including the state of data availability. Chief among these was the question of baselining: without a defined ideal state of resilience or comparative cross-country scores, it is difficult to assess whether Australia’s performance indicates adequate resilience to AI risks.  However, many of the data sources we relied on were themselves index rankings, which imparted a relativistic character to our analysis. 

Data was obtained from government publications, international indices managed by the OECD and civil society, and academic studies. For bio-related risks, the Global Health Security Index \cite{ghsi} proved useful, given that its sub-indicators are transparently reported. By contrast, no cybersecurity index with useful sub-indicator data was available. Obtaining up-to-date information proved challenging in many cases. 

Nonetheless, several opportunities for improving Australia's resilience to advanced AI risks emerged. Australia may benefit from strengthening stockpiling and supply chain readiness for health emergencies, implementing chem-bio procurement screening and preparedness training, improving adherence to cybersecurity best practices across government, and addressing gaps in cyber workforce capacity, especially the proportion of entities meeting core requirements.

\section{Discussion}
\label{sec:discussion}
\subsection{SCAF's Potential Use Cases}
\textit{\textbf{Informing Pre-Deployment Risk Management}}

By offering a systemic view of societal contexts, SCAF could complement existing frontier AI safety frameworks, helping to guide risk thresholds, mitigations, and deployment decisions. For example, SCAF could support AI developers in calibrating and prioritising AI safeguards based on the vulnerabilities and preparedness of different jurisdictions. Those crafting governance frameworks could use SCAF to help prioritise risks and threat models. For example, if a country is meaningfully resilient to cyber risks, an AI system that may possess some offensive cyber capabilities may not attract the same risk prioritisation. Conversely, if SCAF shows cyber resilience is low, this might inform pre-deployment decisions, including the choice of model evaluations, and safeguards, e.g., setting refusal thresholds.

\textit{\textbf{Public Policy Planning}}

SCAF could support planning and policymaking in several ways. It could inform reviews of national preparedness for advanced AI, guiding investments that strengthen societal resilience amid AI diffusion. It could clarify opportunities for resilience-building and support targeted policy interventions that are not AI-specific, while also highlighting data gaps that warrant further research. It could also guide funding priorities and highlight where collaboration across agencies and disciplines is needed. In these ways, SCAF could reinforce the systemic governance of advanced AI. 

\subsection{Advancing SCAF and Future Research}
Stakeholders can refine and expand SCAF's conceptual framework, test its application, and share and validate indicators against each risk category. In addition, future research should: broaden SCAF's risk coverage; develop a robust evidence base for SCAF; and explore how to involve affected communities in SCAF's design and implementation. 

\textit{\textbf{Broadening SCAF's Risk Coverage}}

SCAF should encompass a broader range of risks than the two we selected for illustrative purposes, which were drawn from the International AI Safety Report \cite{bengio2025international} and frontier safety frameworks \cite{metr2025common}. We considered including a category for risks associated with autonomy and potential loss of control over advanced AI, which are represented in some frontier safety frameworks. However, definitions that enable societal capacity assessment remain underdeveloped. Early work has begun to outline measures of human oversight \cite{cihon2024chilling,cihon2025measuring,paiagents}, internal deployments at frontier labs for research and development purposes \cite{stix2025ai,internaldeployment}, and emergency responses to loss-of-control scenarios \cite{somani2025strengthening}. As this literature matures, related risks could be incorporated into SCAF. Future work could also extend SCAF beyond frontier safety risks to encompass other challenges, such as chronic over-reliance on AI systems and associated effects like emotional dependence \cite{pentina2023exploring} and cognitive de-skilling \cite{lee2025impact,bastani2024generative}. It could be further  adapted to assess vulnerability and resilience to sector-specific risks, including AI failures in financial markets.

\textit{\textbf{Developing the Evidence Base}}

Our attempt to apply SCAF to Australia (\hyperref[appendixa]{Appendix A}) illustrated challenges in data availability. Multiple categories of information are relevant to operationalising and improving SCAF. For example:
\begin{itemize}
    \item \textbf{Scoping SCAF} requires evidence of the likelihood and severity of AI risks to determine which risks should be included in, or prioritised by, SCAF. 
    \item \textbf{Applying SCAF} requires indicators for a given country’s vulnerability, coping capacity, and adaptive capacity in response to those risks, as well as timely, high-quality data to measure those indicators. 
    \item \textbf{Refining and selecting stronger indicators for SCAF} requires evidence of causal mechanisms and how AI harms arise in practice, enabling clearer insight into the resources, measures, and policies societies should deploy to cope and adapt effectively.
\end{itemize}

SCAF’s scope should be grounded in emerging research on which risks are most likely and severe, drawing on data about AI adoption trends, use cases, incident prevalence, and market and technology dynamics. It should be iteratively improved as this evidence base matures. Generating this information will, for example, require post-deployment and incident monitoring mechanisms \cite{stein2024postdeployment}, tracking AI-tool use (e.g., proxied by public Model Context Protocol servers \cite{mcp}), and longitudinal AI-human studies to better understand how AI is being used in real-world contexts. 

Developing a robust information base requires collaboration and coordination among multiple stakeholders, including AI developers, researchers and governments. For example, while governments and regulators are well positioned to collect data on regional or sectoral preparedness, AI developers have better visibility into emerging AI capabilities and, through aggregated privacy-preserving usage studies, how they are being used. Bridging these and other complementary perspectives is essential to building a robust evidence base for SCAF and broader AI governance objectives.

\textit{\textbf{The Role of Participatory Governance}}

Berman et al. (\citeyear{berman2024troubling}) critique attempts to exhaustively taxonomise the potential societal impacts of generative AI. They observe that such attempts invite ‘comparison, and trade-offs, across disparate categories...' and implicitly frame societal impacts as modular, independent, and commensurable. Developers are positioned as arbiters of which impacts to prioritise and how to address them, even though balancing competing impacts is a value-laden task that requires the input and leadership of affected communities. Berman et al. (\citeyear{berman2024troubling}) emphasise that who decides which impacts (or risks) to focus on matters. Although SCAF is neither a taxonomy nor an exhaustive schema of potential AI impacts, we are wary of Berman et al.'s (\citeyear{berman2024troubling}) critique. SCAF's further development and operationalisation could draw inspiration from their call for a participatory-first approach. This would entail accountability to, and participation of, affected communities to enable deliberation over which risks to include and prioritise. 

\section{Conclusion}
Drawing on societal resilience literature, SCAF offers a way to align risk management of advanced AI with its societal deployment context. The indicators-based framework can be used to support multiple stakeholders in assessing and strengthening societal resilience ahead of widespread AI diffusion. SCAF demonstrates an initial step towards more informed risk management of advanced AI. To be strengthened, SCAF will require clearer scoping of which risks to include and prioritise, stronger indicators grounded in evidence of causal mechanisms, and timely, high-quality data to measure those indicators. As the evidence base matures—through post-deployment monitoring, incident tracking, longitudinal studies, and related research—there is an opportunity to build a richer understanding of how advanced AI interacts with societies, and what factors shape their vulnerability and capacity to cope and adapt. Finally, SCAF's legitimacy depends on participatory processes that bring affected communities into decisions about which risks matter and how they should be addressed. By working together on these shared challenges, we can develop applied methods to assess and strengthen societal resilience to advanced AI.

\section*{Acknowledgements}
Thanks to Seth Lazar, Melissa Parsons, Samantha Copeland, Erica van Ash, Risto Uuk, Brandon Jackson, Shahar Avin, Andrew Strait, Amanda Craig Deckard, Hector de Rivoire, Simon Staffell, Mike Linksvayer, and TAIG ICML 2025 workshop participants. Any errors are the authors' alone.


\section*{Impact Statement}

This paper advances risk assessment and risk management for advanced AI. SCAF may be applicable to many risks, though we focus narrowly on two specific risks. Please see Section 4 for additional context.


\bibliography{example_paper}
\bibliographystyle{icml2025}

\newpage
\appendix
\onecolumn

\section{Prototype SCAF with Indicators}
\label{appendixa}
To support the translation of SCAF from theory to practice, Table \ref{tab:scaf-prototype} presents a prototype set of relevant indicators, though its scope remains limited in terms of both the risks and capitals considered. Table \ref{tab:scaf-australia} applies this prototype to the case of Australia.

\subsection{Australia Case: Background}

We selected Australia because of our familiarity with its AI public policy landscape and the expectation that relevant data would be publicly accessible. Adoption of AI is steadily increasing: by April 2025, 44 percent of small and medium enterprises reported using or intending to use AI, up from 35 percent a year earlier \cite{ausAdoption}; in 2024, 1.14 percent of all job postings were AI-related, ranking Australia 14th globally \cite{maslej2025artificialintelligenceindexreport}; and 50 percent of Australians reported semi-regular or regular AI use, which is lower than the average (66 percent) across 47 countries \cite{kpmg}. Cloudflare Radar data suggests that bot activity in Australia is moderately lower than in other regions \cite{cloudflareAus}. 

In terms of public governance of AI risks, Australia lacks binding horizontal AI regulation, and its policy discourse has shifted from a safety-first orientation to emphasising capability and adoption, particularly for productivity gains \cite{barlow2025labor}. In August 2024, the Department of Industry, Science and Resources published the Voluntary AI Safety Standard based on frameworks such as ISO/IEC 42001:2023 and NIST's AI Risk Management Framework \cite{aiStandard}. We could not find documented government use of AI incident databases, which could suggest a lack of systematic visibility of AI-related failures.

Australia is not a frontier developer of general-purpose AI models \cite{maslej2025artificialintelligenceindexreport}, though it has perceived strengths in research and applied AI in domains such as mining, healthcare, and agriculture \cite{bratanova2025ecosystem}. Australian society and institutions are intertwined with foreign technology platforms across social media, software, and cloud computing, which may loosely predict Australia’s exposure to AI-related disruptions. For example, in 2021, Facebook’s eight-day ban on Australian news content disrupted public health communications during the COVID-19 pandemic \cite{bruns2021biggamble}, while in July 2024 a faulty software update from U.S.-based CrowdStrike caused a global ICT outage that severely disrupted Australian banks, airlines, police, media, and telecommunications, resulting in losses exceeding \$1 billion AUD \cite{daly2024crowdstrike,ongweso2024msftcrowd}.

\subsection{Australia Case: Findings}

We analysed data on Australia’s vulnerability, coping capacity, and adaptive capacity in relation to the risk that advanced AI will be misused to conduct cyber, chemical and biologial attacks. Identifying relevant data proved challenging. Moreover, the inferences we could draw from the available data were difficult to interpret without a baseline. This limitation arises because we have neither defined an ideal state of resilience nor collected cross-country data that would allow Australia’s performance to be assessed relative to other countries. Nonetheless, many of the sources we identified were themselves index rankings of Australia, which imparted a relativistic character to our attempted application of SCAF. 

Ultimately, the the strongest findings we can draw from this analysis are that Australia may have opportunties for improving:
\begin{itemize}
    \item stockpiling and supply chain readiness for health emergencies;
    \item implementing chem-bio procurement screening and preparedness training;
    \item     adherence to cybersecurity best practices across government; and
    \item cyber workforce capacity, particularly the proportion of entities meeting core cybersecurity requirements. 
\end{itemize}
The implication is that making these improvements may be important to Australia's preparedness for emerging AI capabilities.
\begin{table*}[htbp]
\centering
\caption{Prototype SCAF with Indicators}
\footnotesize
\begin{tabularx}{\textwidth}{|>{\raggedright\arraybackslash}p{0.15\textwidth}|>{\raggedright\arraybackslash}X|>{\raggedright\arraybackslash}X|}
\hline
 & \textbf{Chemical and biological (CB) risks} & \textbf{Cyber risks} \\
\hline
\textbf{Vulnerability} 
& 
Capacity of public healthcare systems (e.g., proxied by hospital bed per capita) \cite{keegan2013measuring,grima2020country,thomas2020strengthening,george2021critical,fleming2022metrics}.
\vspace{4pt}

Population health status (e.g., proxied by infant mortality rate) \cite{grima2020country,stockwell1987trends}.
\vspace{4pt}

Population flows and density, including frequency of mass gathering \cite{grima2020country,calka2017fine}.
\vspace{4pt}

Presence of extremist and terrorist groups and extent of their activities \cite{caskey2013global}.
\vspace{4pt}

Available CB weapons materials, expertise and equipment \cite{caskey2013global}.
& 
Societal reliance on digital services (e.g., proxied by internet penetration rates and internet banking rates). 
\vspace{4pt}

Quantity and severity of cyber attacks (e.g., proxied by regional financial risk exposure of cyber risk type and share of enterprises that experienced cyber incidents) \cite{oecd2024new}.

\\[10pt] 
\cline{2-3}
& \multicolumn{2}{>{\raggedright\arraybackslash}p{0.80\textwidth}|}{
\textbf{Cross-cutting vulnerability indicators}:
\vspace{4pt}

Economic development (e.g., proxied by GNI per capita, GDP, population below \$2.15 per day) \cite{grima2020country,dynan2018gdp}.
\vspace{4pt}

State capacity (e.g., proxied by tax revenue as a share of GDP, government territorial control, skills and impartiality of public service) \cite{herre2023state}.
\vspace{4pt}

Democracy and peace \cite{bolpagni2022cyber}.
}\\
\hline
\textbf{Coping Capacity} 
& 
Health emergency response policies and plans. 

\vspace{4pt}

Capacity of emergency response services. 

\vspace{4pt}

Regional crisis management system. 

\vspace{4pt}

Performance standards for CB protective equipment. 

\vspace{4pt}

CB surveillance and reporting system. 

\vspace{4pt}

Medical countermeasures stockpiles. 

\vspace{4pt}

CB response training for first responders and medical professionals. 

\vspace{4pt}

CB specialists (e.g., no. of NATO ICBRN-FR / NFPA CBRN / equivalent qualifications). 

& 
Cyber incident tracking and response, including in critical infrastructure, government, and other sectors. 

\vspace{4pt}

Regional cyber response coordination system.

\vspace{4pt}

Number of cybersecurity security professionals (e.g., number of GitHub developers \cite{github2025innovation}). 

\vspace{4pt}

Patch adoption rates.
\\
\hline
\textbf{Adaptive Capacity}
& 
Medium to long-term health emergency preparedness policies and planning. 
\vspace{4pt}

CB material procurement standards and monitoring (e.g., synthetic nucleic acid procurement screening).
\vspace{4pt}

CB monitoring and threat sharing. 
\vspace{4pt}

Implementation of NATO CB defense policy / equivalent.
\vspace{4pt}

Critical infrastructure protections for public health system.
\vspace{4pt}

Citizen preparedness. 
\vspace{4pt}

\textit{Transformative Capacity Examples: }
\vspace{4pt}

AI-enabled patching for expanded synthetic nucleic acid procurement screening.  
\vspace{4pt}

AI-enabled biological monitoring and detection. 
\vspace{4pt}

Development of AI forecasting and detection tools for multihazard risk management. 
\vspace{4pt}

AI-enabled early warning systems. 
\vspace{4pt}

AI-enabled counterterrorism tools. 

& 
Medium to long-term cybersecurity talent pipeline.
\vspace{4pt}

Secure-by-design initiatives. 
\vspace{4pt}

Adoption of cybersecurity best practices by local websites and open-source developers.
\vspace{4pt}

Local implementation of NIS-2 / equivalent critical infrastructure cybersecurity measures. 
\vspace{4pt}

Critical infrastructure protection targeted at cyber resilience. 
\vspace{4pt}

Cross-sector security standards. 
\vspace{4pt}

AI literacy. 
\vspace{4pt}

Cyber hygiene literacy. 
\vspace{4pt}

\textit{Transformative Capacity Examples:}
\vspace{4pt}

Active cyber defence policies and measures. 
\vspace{4pt}

Government and critical infrastructure adoption of AI tools for cybersecurity, including detecting and mediating vulnerabilities at scale. 
\vspace{4pt}

AI cyber threat monitoring. 
\vspace{4pt}

AI-enabled cyber defense. 

\\
\hline
\end{tabularx}
\label{tab:scaf-prototype}
\end{table*}

\begin{table*}[htbp]
\centering
\caption{Illustrative SCAF: Australia}
\scriptsize

\begin{tabularx}{\textwidth}{|>{\raggedright\arraybackslash}p{0.15\textwidth}|>{\raggedright\arraybackslash}X|>{\raggedright\arraybackslash}X|}
\hline
 & \textbf{Chemical and biological (CB) risks} & \textbf{Cyber risks} \\
\hline
\textbf{Vulnerability} 
& 
\textit{Healthcare system capacity:} 
\vspace{0pt}

Global Health Security Index (GHSI) score: 72.2\% (5th globally) \cite{ghsi}; hospital beds: 2.50 per 1000 population \cite{australiaHospital} vs. 4.6 OECD average \cite{oecdHospital}.
\vspace{4pt}

\textit{Population health status:} 
\vspace{0pt}

 GHSI 83\% (6th globally) \cite{ghsi}; infant mortality: 3.2 per 1000 live births \cite{australiainfant} vs. 4.0 OECD average \cite{OECDinfant}. 

& 
\textit{Cyberattack incidence:}
\vspace{0pt}

30.5\% of businesses with 10 or more employees vs. OECD average of 19.53\% \cite{oecdICTbreach}. Publicly reported common vulnerabilities and exposures increased 31\% in 2023, with additional stats on incidents, critical infrastructure notifications, and cybercrime costs \cite{australiaCyber}.
\vspace{4pt}

\textit{Reliance on digital infrastructure:}
\vspace{0pt}

Internet penetration rate: 97.1\% \cite{australiaITU}.

\\[10pt] 
\cline{2-3}
& \multicolumn{2}{>{\raggedright\arraybackslash}p{0.80\textwidth}|}{
\textbf{Cross-cutting vulnerability indicators}:
\vspace{4pt}

\textit{Political and security risk} (including government effectiveness, risks of social unrest, illicit activities by non-state actors, terrorist attack risks, territorial control, and armed conflict): GHSI 80.1\% (31st globally) \cite{ghsi}.
\vspace{4pt}

\textit{Economic development}: (GNI: \$69454 USD/capita, 10th highest in OECD \cite{gniAus}. GDP: 1.75T USD, 13th globally \cite{worldbankAustralia}; 0.5\% of population living below \$3.00 per day \cite{worldbankAustralia}.
}
\\
\hline
\textbf{Coping Capacity} & 
\vspace{0pt}

\textit{Health emergency response policies and plans:}
\vspace{0pt}

GHSI overall biosecurity 62.7\% (9th), 66.7\% (15th) for emergency preparedness and response planning \cite{ghsi}.
\vspace{4pt}

\textit{Surveillance and reporting system:}
\vspace{0pt}

GHSI 82.2\% (2nd), including 100\% (1st) for real time surveillance and reporting \cite{ghsi}.
\vspace{4pt}

\textit{Supply chain for health system (including countermeasure stockpiles):}
\vspace{0pt}

GHSI 61.1\% (15th) \cite{ghsi}; yet Schdmidt \& Schdmit, (\citeyear{schmidt2024chemical}) raise concerns that countermeasure stockpiles are 'generally too small'. 
\vspace{4pt}

GHSI: 100\% for risk communication \cite{ghsi}; further survey data available on psychological and material preparedness \cite{every2019australian}.
\vspace{4pt}

\textit{CB response training:}
\vspace{0pt}

No CB health course since 2012 in Joint Health Command of Australian Defense Force \cite{heslop2019medical}; anecdotal: 2/6 hospitals in Queensland hospital survey conducted CB training in past 12 months \cite{mackie2022chemical}; more general epidemiology workforce scores 100\% under GHSI based on a measure of at least 1 field epidemiologist per 200000 people \cite{ghsi}.

& 

\textit{Cybersecurity professionals:}
\vspace{0pt}

Proxy: open source developers on GitHub: 1.8 million as of 2025 (19th globally) \cite{github2025innovation}.
\vspace{4pt}

\textit{Emergency response:}
\vspace{0pt}

The Australian Cyber Security Centre (ACSC) serves as the national Computer Emergency Response Team (CERT Australia) and 24/7 cyber incident response hub. It coordinates technical incident response and advice nationwide as the lead operational agency for cyber emergencies. 

\vspace{4pt}
Also see \href{https://www.homeaffairs.gov.au/cyber-security-subsite/files/australian-cyber-response-plan.pdf}{Australia's Cyber Response Plan} (June 2025), the \href{https://www.homeaffairs.gov.au/about-us/our-portfolios/cyber-security/cyber-coordinator/cyber-security-response-coordination-unit}{cyber security response coordination unit}, and the \href{https://www.pmc.gov.au/resources/australian-government-crisis-management-framework-agcmf}{Australian Government Crisis Management Framework}. 

\textit{Monitoring and detection:}
\vspace{0pt}
The ACSC operates a two-way \href{https://www.cyber.gov.au/about-us/view-all-content/news-and-media/join-the-cyber-threat-intelligence-sharing-service-through-sentinel}{Cyber Threat Intelligence Sharing (CTIS) platform} that enables government and industry partners to rapidly share and receive cyber threat data in real time (at machine speed). 
\vspace{4pt}

\\
\hline
\textbf{Adaptive Capacity}
& 
\textit{CB procurement and monitoring:}
\vspace{0pt}

GHSI 50\% (2nd) for for dual-use research and culture of responsible science, notably finding that Australia does not have screening requirements for synthesised DNA prior to sale \cite{ghsi}.

\vspace{4pt}

\textit{NATO CBRN defence policy or equivalent:}
\vspace{0pt}

Domestic Health Response Plan for CBRN Incidents of National Significance \citeyear{australiaCBRN}, Assessments: COVID revealed gaps; \cite{mackie2022chemical} found anecdotal deficiencies. \href{https://docs.google.com/document/d/1hXFxm74lup2pLku0Zoo2hBEg3m7wzaHFc8g_WDa1qZ8/edit?usp=sharing}{CLAUDE grades B-} vs. NATO standard \cite{NATOcbrn}.

\vspace{4pt}

\textit{Citizen preparedness:}
\vspace{0pt}

GHSI: 100\% for risk communication \cite{ghsi}; further survey data available on psychological and material preparedness \cite{every2019australian}. 

\vspace{4pt}

\textit{Transformative capacity:}
\vspace{0pt}

Embed new technologies in biosecurity screening, traceability, and response. Identified at a high-level in the 2022-2032 National Biosecurity Strategy \cite{biosecurity2022}. Increased ambition could include piloting AI-informed screening patches \cite{wittmann2024toward}. 

& 

\textit{Best practice utilisation:}

\vspace{0pt}

70\% of businesses report implementing preventative measures to address cybersecurity threats \cite{australianStat}.

\vspace{0pt}

ITU Global Cybersecurity Index \cite{ituCyber} ranks Australia in Tier 1 "Rolemodeling" globally.

\vspace{0pt}

Only 15\% of government entities met required cybersecurity practices in 2024 \cite{australianCyberPosture}.

\vspace{4pt}

\textit{Secure-by-design initiatives: }
\vspace{0pt}

\href{https://www.cyber.gov.au/resources-business-and-government/governance-and-user-education/secure-by-design/secure-design-foundations}{Australia: Secure by Design foundation}.

\vspace{0pt}

\href{https://www.cyber.gov.au/about-us/news/asdacscncsc-collaboration-strengthen-cyber-security-posture}{Cyber Hygiene Improvement Program} for governments.

\vspace{4pt}

\textit{Active cyber defense initiatives:}

\vspace{0pt}

\href{https://www.qld.gov.au/community/your-home-community/cyber-security/cyber-security-for-queenslanders/report-a-cyber-crime}{Australian Cyber Security Hotline}.

\vspace{4pt}
\textit{Transformative capacity:}
\vspace{0pt}

ACSC partnered with Microsoft to develop a plugin for Sentinel that enables businesses to contribute to the CTIS platform. 
\vspace{0pt}
Implementation of 2023-2030 Australian Cyber Security Strategy, particularly 12(1) next-generation threat blocking capabilities \cite{cyberstrategy}.

\\
\hline
\end{tabularx}

\label{tab:scaf-australia}
\end{table*}


\end{document}